# Engineering Spin Coherence in Core-Shell Diamond Nanocrystals

Uri Zvi[1], Denis R. Candido[2], Adam Weiss[3], Aidan R. Jones[4], Lingjie Chen[1,4], Iryna Golovina[3,5], Xiaofei Yu[4], Stella Wang[4], Dmitri V. Talapin[1,3,6], Michael E. Flatté[2,7], Aaron P. Esser-Kahn[1], Peter C. Maurer*[1,6]

[1]Pritzker School of Molecular Engineering, University of Chicago, Chicago, IL 60637, USA.

[2]Department of Physics and Astronomy, University of Iowa, Iowa City, Iowa 52242, USA.

[3]The Department of Chemistry, University of Chicago, Chicago, IL 60637, USA.

[4]The Department of Physics, University of Chicago, Chicago, IL 60637, USA.

[5]The Laboratory for Research on the Structure of Matter, University of Pennsylvania, Philadelphia, PA 19104, USA.

[6]Center for Molecular Engineering and Materials Science Division, Argonne National Laboratory, Lemont, IL, USA.

[7]Department of Applied Physics, Eindhoven University of Technology, Eindhoven, 5600 MB, The Netherlands.

*pmaurer@uchicago.edu

**Diamond nanocrystals can harbor spin qubit sensors capable of probing the physical properties of biological systems with nanoscale spatial resolution[1]. These diamond nanosensors can readily be delivered into intact cells[2] and even living organisms[3]. However, applications beyond current proof-of-principle experiments require a substantial increase in sensitivity, which is generally limited by surface-noise-induced spin dephasing and relaxation[4]. In this work, we significantly reduce magnetic surface noise by engineering core-shell structures, which in combination with dynamical decoupling result in qubit coherence times ($T_2$) ranging from 52μs to 87μs – a drastic improvement over the 1.1μs to 35μs seen in bare particles. This improvement in spin coherence, combined with an overall increase in particle fluorescence, corresponds to a two-order-of-magnitude reduction in integration time. Probing qubit dynamics at a single particle level, furthermore, reveals that the noise characteristics fundamentally change from a bath with spins that rearrange their spatial configuration during the course of an experiment to a more dilute static bath. The observed results shed light on the underlying mechanisms governing spin dephasing in diamond nanocrystals and offer an effective noise mitigation strategy based on engineered core-shell structures.**

Nitrogen vacancy (NV) centers in diamond nanocrystals have emerged as a powerful platform for sensing magnetic fields[5,6], electric fields[7], and temperature[8,9] in living systems. First applications of diamond-based sensing are emerging in neuroscience[10], developmental biology[9], cellular physiology[11,12], and medical diagnostics[13]. However, many potential applications relying on diamond nanocrystal-based sensing remain limited by relatively short NV $T_2$-times, which are typically 100x shorter than what is observed in high-purity bulk crystals[14]. Specifically for particles with a size below 100nm, charge and spin noise associated with the crystal's surface become a dominant factor[4]. Various approaches to extending the spin coherence in diamond micro- and nanocrystals have been pursued. Notably, these approaches include mechanical milling[15,16] or lithographically[17] etching high-purity bulk diamond to microscale particles, which under dynamical decoupling, results in bulk-like $T_2$-times. Although promising, the microscale size (hundreds of nanometers[15,16] in the case of milling and pillar length of a few micrometers[17] in the case of lithographic etching) of these particles severely limits biological applications. Moreover, the top-down fabrication required in lithography results in a low yield, limited to tens of micrograms, which makes processing large quantities prohibitively expensive. Another approach towards the mitigation of surface noise relies on controlling the diamond surface termination, which on highly ordered bulk diamond surfaces, has led to a significant increase in spin coherence[18]. However, in diamond nanocrystals a similar effect on coherence has yet to be observed.

In nanotechnology, engineered core-shell structures have been established as an effective strategy to mitigate adverse surface effects on luminescence. For example, encapsulation in a protective shell reduces surface-induced photoblinking in quantum dots[19] and non-radiative relaxation in lanthanide-doped upconverting nanoparticles[20,21]. In principle, a similar coating strategy could lead to enhanced qubit coherence by saturating dangling bonds and eliminating paramagnetic defects or charge traps located near the particle's surface[21]. However, extending qubit coherence in core-shell structured particles has so far remained elusive[4,22]. In this work, we explore core-shell structures to efficiently passivate the diamond surface. This leads to a $T_2$ extension in diamond nanocrystals that rivals those of near-surface NVs in high-purity bulk crystals[18]. As a substrate, we use electron irradiated and thermally annealed (850 °C) carboxylated diamond nanocrystals with a reported average diameter of 44nm and 10-12 NV centers per particle[23]. The diamond nanocrystals were coated with $18\pm2$nm silica shells using an adapted Stöber process with tetraethyl orthosilicate (TEOS) to obtain core-shell structured particles[24], resulting in milligrams of diamond core-shell particles (Fig. E1, SI1).

Transmission electron microscopy (TEM) reveals dense silica shells with homogeneous surface coating for these core-shell structures (Fig. 1A and E2). Correlative light-electron microscopy (CLEM) imaging reveals that core-shell structures result in a 1.85-fold increased luminescence (see methods) for a given diamond core size (Fig. 1B, E3, and SI2-3). Fluorescence spectroscopy confirms that our core-shell structures stabilize the desired negatively charged NV center (Figure 1C). Deconvolution of the fluorescence spectra of $NV^0$ and $NV^-$ suggests a 20% increase of $NV^-$. Silica has been reported to efficiently reduce surface states[25] that lead to a decrease in charge stability and a quenching of the fluorescence signal[26,27]. An additional fluorescence increase can be attributed to a larger photonic density of state in the silica shell when compared with air[28].

We start by investigating the type and density of paramagnetic defects present in bare and core-shell structured diamond nanocrystals. Continuous wave electron paramagnetic resonance (EPR) spectroscopy on lyophilized bare and core-shell structured nanocrystals reveals the spectroscopic signatures of at least three distinct resonances (Fig. 2A,B). One resonance can be attributed to the presence of substitutional nitrogen defects (P1 centers) identified by their characteristic hyperfine interactions with the $^{14}$N nuclear spin[29]. The second resonance, henceforth referred to as X-spins, has been suggested to correspond either to dangling bonds[30] or negatively charged vacancies in the near-surface region[31] – for both defects, the expected g-factor lies within the precision of our spectrometer. The remaining third resonance can be assigned to a hydrogen atom-vacancy (H1) complex[32].

Remarkably, the core-shell structures show a significant reduction of all resonances when compared to the bare diamond nanocrystals (i.e., X-spin density is reduced by 3.8x, H1 by 2.6x, and P1 by 1.8x). The depletion in X and P1 defects points toward a band-bending at the diamond-silica interface due to changes in surface potential[33]. We illustrate this effect by aligning the electronic structures of type Ib oxygen-terminated diamond (N doped at ~100 ppm) and amorphous silica (Fig. 2C and Fig. E5 for band structure and Fig. 2D for a schematic representation). The reported surface electron affinity, $\chi_S \sim 2eV$, of oxygen-terminated bare diamond[34] leads to a downward band-bending that stabilizes P1 and X-spins (left panel). In contrast to bare diamond, silica encapsulation results in an upwards band bending, which depletes P1 and X-spins without affecting the energetically lower laying NV$^-$ charge state (right panel). In addition to the depletion of noisy paramagnetic species, the large energy barrier (~1 eV) prevents tunneling of electrons deep into the SiO$_2$[35], which is expected to result in an increased charge stability during photoexcitation. We confirm our model with a band bending simulation based on a solution of the Poisson equation (see methods, Fig. E6 and SI6). The resulting band structure suggests a 3.8nm thick P1 depletion layer at the diamond-silica interface, which translates into a 44% decrease in the number of P1 centers per nanoparticle. As supported by our band bending model, our simulation predicts that the NV$^-$ density remains largely unaffected by our core-shell structures. We note that direct chemical conjugation of dangling bonds or displacement of paramagnetic species in the hydrolyzation layer by silica encapsulation can also result in a reduction of paramagnetic spins at the surface[36].

EPR spectroscopy provides important insights into the presence and density of paramagnetic defects in bare and core-shell particles. However, ensemble EPR spectroscopy does not account for particle heterogeneity and, at least in our case, does not possess the sensitivity to directly probe NV-qubits (NV centers have a ~100x reduced density compared to P1 centers[23]). We overcome this challenge by selectively probing the NV spins within individual optically resolvable diamond nanocrystals. Double electron-electron resonance (DEER) measurements confirm the coupling of the NV-qubit to X-spin and P1 (Fig. E7 and SI7). Figure 3A shows the observed $T_1$-times for bare and core-shell diamond nanocrystals drop-casted on a glass substrate. The single quantum relaxation time ($T_1^{SQ}$), which describes relaxations between $m_s = 0$ and $m_s = \pm 1$, sharply increases from $114 \pm 16 \mu s$ for uncoated particles to $379 \pm 33 \mu s$ for engineered core-shell structures. However, double quantum relaxations ($T_1^{DQ}$), transitions between $m_s = \pm 1$, remain unaffected by coating (inset, Figure 3A). This suggests that low-frequency electronic noise is not impacted by our engineered core-shell structures. Using CLEM measurements, we confirm that the increased NV $T_1$-times are not the result of selecting core-shell structures with an increased core diameter, but rather the consequence of the engineered material properties. Figure 3B shows $T_1^{SQ}$ and core size for four bare and four core-shell particles as extracted by CLEM (Fig E3 and SI3). Although the fluorescent particles are nominally 44nm in diameter, we observe that particles with obtainable spin coherence possess diameters of 70nm. We point out that different characterization techniques can result in significantly different particle size estimations. For example, atomic force microscopy (AFM) measures the particle height[15] whereas TEM measures the particle cross-section. In the case of disk-like particles, such as diamond nanocrystals produced by ball milling, AFM will, therefore, consistently underestimate the particle size when compared with TEM[37].

Having established that engineered core-shell structures result in an increase in qubit $T_1$, we next investigate its effect on $T_2$. From spin Echo experiments, we find that bare and coated particles have a $T_2^{Echo}$ of $1.05\pm0.41\mu s$ and $1.42\pm0.20\mu s$, respectively (Fig. E8 panels C and E). Using Carr-Purcell-Meiboom-Gill (CPMG) dynamical decoupling[38] we extend the coherence by filtering low-frequency noise with a filter function that is centered around $\omega = \pi N/T$, where $N$ is the number of π-pulses and $T$ is the total precession time[38]. We find that for bare nanocrystals, $T_2$ increases with $N$ only up to a certain level before it saturates (Fig. E8 panels A and F). Figure 3B shows the maximally achievable $T_2$ for twelve different bare diamond particles, with some particles showing no or marginal improvement with $N$ (blue data points). Interestingly, for core-shell structured nanoparticles (green data points) we do not observe a similar heterogeneity in $T_2$ and find a significant $T_2$ increase for all eight particles (for $N > 1,000$ we find an average $T_2^{CPMG} = 70\pm12\mu s$). As a consequence core-shell structures result in a average 3.5-fold increase in qubit coherence, and a 3.2-fold decrease in particle to particle $T_2$ variation (i.e., the relative standard deviation is $\frac{\sigma(T_2)}{\langle T_2 \rangle} = 0.60$ for bare and $\frac{\sigma(T_2)}{\langle T_2 \rangle} = 0.19$ for core-shell particles). The functional dependence of $T_2$ on $N$ follows a power law, $T_2(N) = T_{2,echo} N^k$, with $k = 0.53$ for core-shell and $k = 0$ to $0.47$ for bare particles. The difference in scaling suggests that the spin-bath noise in core-shell structured particles is characterized by longer correlation times compared to bare particles[38] (SI9). Again, CLEM measurements (Fig. 3D) of four bare and four core-shell nanoparticles confirm that the cores of the studied particles are of comparable sizes.

The different performance under CPMG decoupling points toward a modification of the spin noise environment in our core-shell structured particles. Assuming Gaussian noise, we reconstruct the power spectra by deconvoluting the experimentally measured CPMG time traces (see reference[38] and methods), which results in the solid circles in Figure 4A. In addition to the frequency range obtained from CPMG (2MHz to 25MHz), we can probe high-frequency noise ($\sim 2.85$GHz) by considering the measured $T_1^{SQ}$ spin relaxation times (circle in Fig. 4A). In the low-frequency regime (<3.7MHz), we find that the noise power spectra of bare and core-shell structured nanocrystals are largely identical, while for higher frequencies the engineered core-shell structures show noise reduction of up to a factor $4.0$x.

The power spectrum of a spin bath follows a Lorentzian[38]. For bare particle we observe a broad power spectrum that fits a Lorentzian with short correlation times ($\tau_c \leq 1ns$), as would be expected for a fast fluctuating surface spin bath[39]. In contrast, for core-shell particles this high-frequency noise is significantly reduced, revealing a Lorentzian noise spectrum with longer correlations times ($\tau_c = 46 \pm 11ns$), indicating a slower evolving spin bath. At low frequencies, the power spectrum deviates from a Lorentzian noise model and instead follows a $1/f$-like scaling. The exponent ($a = 1.7$ and $a = 1.6$ for bare and core-shell particles, respectively) of this $1/f^a$-noise is extrapolated from the experimentally observed $T_1^{DQ}$ and is in good agreement with results from near-surface NV centers in bulk diamond[40]. This suggests that charge, rather than spin, fluctuations dominate the low-frequency end of the noise power spectrum[4] (see methods and Fig. E9) and remain unaffected by our engineered core-shell structures.

Having gained an understanding of the noise spectral properties, we next turned our attention to the microscopic origin and quantum mechanical properties of the spin bath. In a Hahn Echo, the coherence factor $\exp(-\chi(t))$ follows a stretched exponential with $\chi(t) \sim t^n$, where the exponent $n$ implicitly contains information about the spin bath[41]. Figure 4B shows four representative examples of the time evolution of $\chi(t)$ for bare and core-shell structured particles. In the case of core-shell particles we find $n \sim 1$, which is consistent with a Markovian bath where the spatial position of each bath spin remains fixed[41] (Figure 4C and SI11). In contrast, Hahn Echo for bare nanocrystals shows a strikingly different behavior, with $n$ ranging from $0.4$ to $2$ (Figure 4B,C). The observation of $n < 1$ suggests that, for bare particles, the bath spins do not remain in a fixed spatial configuration, but rather change their spatial distribution over time. A similar 'spin-hopping' effect is known to occur in near-surface NV centers in bulk diamond[42,43]. Therefore, the $n \sim 1$ exponent for core-shell particles suggests that the engineered shells not only reduce the paramagnetic defects density, but also reduce spin-hopping within the bath – an effect that has been plaguing diamond-based quantum sensing[42].

Pointing out the exact microscopic identity of the surface-related paramagnetic defects that limit NV coherence in diamond nanocrystals and near-surface bulk systems remains an outstanding challenge. Consistent with our EPR results, recent theoretical work has proposed that surface-related sp$^2$ and sp$^3$ dangling bonds are a major source for NV dephasing[25,44]. We show here that engineered core-shell structures are an efficient way to suppress these surface-related defects and hence alter the spin bath properties, which leads to a significant increase in NV-qubit coherence. The presence of silica-related defects, such as hydroxyl groups, near the diamond-silica interface can furthermore lead to electron trapping[45] and so suppress the rate of spin-hopping, an effect corroborated by modification in the stretch factor of the observed Hahn echo (Fig. 4B,C).

We demonstrate that engineered core-shell structured particles offer an efficient means to reduce heterogeneity and extend qubit coherence. In a central spin model, the qubit coherence is inversely proportional to the spin bath density[46], which puts the observed 3.5-fold increase in NV coherence in good agreement with the 3.88-fold reduction of X-defect density obtained by EPR spectroscopy. Combining our core-shell structures with dynamical decoupling we extended the NV-qubit coherence to 70μs, which approaches the coherence times of near-surface NV centers in high-purity bulk diamond[18]. Furthermore, combining optical single-particle addressability with NV-based qubit sensing reveals that engineered core-shell structures reduce qubit heterogeneity and suppress complex dynamics in spatial reconfigurations of the spin bath during the course of an experiment.

The significantly decreased heterogeneity in core-shell structured particles will be directly applicable to real-world quantum sensing experiments where large variations in $T_1$ and $T_2$, and therefore sensitivity, present a challenge to experimental reproducibility. Biophysical problems ranging from nanoscale thermometry[9] to the detection of paramagnetic species[11] to magnetometry[12] are expected to immediately benefit from such novel noise mitigation strategies. Depending on the measured particle, the observed increase in spin coherence and particle luminescence translates into a 4 to 120-fold reduction in signal integration time for the detection of a phase-coherent signal and a 5 to 1,000-fold reduction for the detection of an incoherent signal [47].

An effective passivation of X-spins and a desired enhancement in coherence might be achieved with thinner shells down to a few nm in thickness. Such a reduction in shell thickness will be central for sensing applications where a minimal spatial separation between the NV-qubit sensor and the target is required. Investigating different coating protocols and chemistry may further extend coherence and at the same time provide deeper insights into the microscopic nature of the X-spins. The developed techniques can directly be extended to other qubit systems, including color centers in silicon carbide[48], lanthanide-doped nanoparticles[21], and quantum dots[22]. Likewise, other diamond nanostructures where surface-induced NV dephasing is a limiting factor, such as in diamond-AFM tips and photonic structures, can benefit from a similar passivation approach. Finally, through standard silanization chemistry[49] the $SiO_2$ surface of our core-shell particles can also serve as the basis for surface functionalization[49], and subsequent targeting of biological molecules and structures with highly coherent qubit sensors in-vivo and in-vitro. Considered together, our findings emphasize the potential of engineering spin coherence using fundamental nanoscience principles to significantly improve the sensitivity of real-world nanoscale quantum sensors.

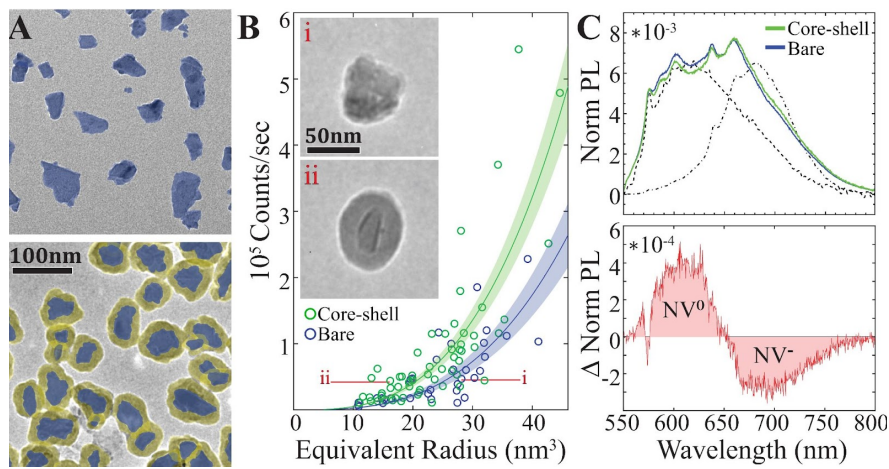

**Figure 1 — Optical properties of bare and core-shell particles. (A)** False-colored TEM of Bare (top panel) and core-shell (lower panel) particles. Blue corresponds to the diamond core and yellow to the silica shell (for raw TEM images see Fig. S2). The aggregation of core-shell particles is an artifact of drying the solution on the TEM grid (see SI). **(B)** CLEM measurements of fluorescence of core-shell (green circles) and bare (blue circles) particles as a function of core size. Solid lines represent a fit to $y = ar^3$, where $r$ is the core radius and $a$ is a fit parameter. Inset: representative TEM of a bare (i) and a core-shell (ii) particle, marked in red on the main panel. **(C)** Normalized spectrum (Top panel) obtained from an ensemble of core-shell (green) and bare (blue) particles. Dashed black lines represent deconvolution to $NV^0$ (left) and $NV^-$ (right) (black lines are extracted from Figure 1 in reference[50]). Subtraction (lower panel) of the core-shell spectrum from bare spectrum showing a shift from $NV^0$ to $NV^-$ in core-shell particles. Spectral decomposition reveals that for bare diamond, 24% of the emission originates from $NV^-$ and for core-shell structures, 29% originates from $NV^-$.

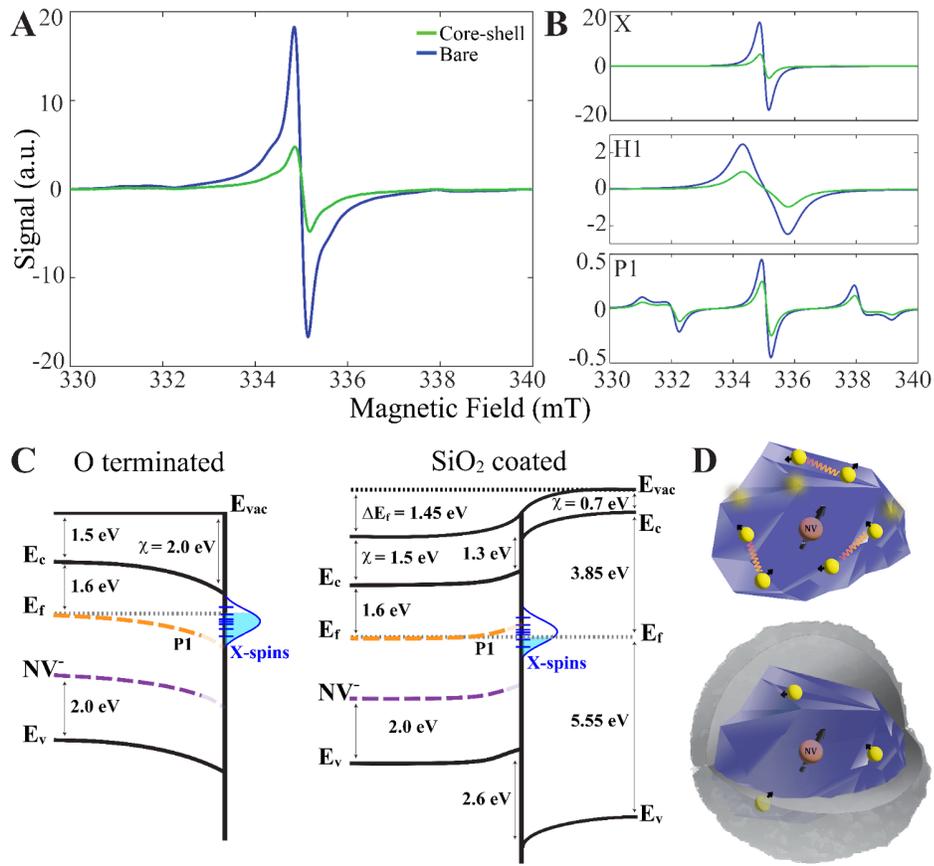

**Figure 2 — EPR of paramagnetic defects in bare and core-shell particles. (A)** Continuous wave X-band EPR ($g = 2.0031 \pm 0.00005$) for bare (blue) and core-shell (green) diamond nanocrystals with ~70nm core size. **(B)** Spectral deconvolution of the EPR signal into the signal from X ($g = 2.0032$), H1 ($g = 2.0028$), and P1-spins ($g = 2.0026$). normalized to a sample mass of 1 mg. All measurements were performed at room temperature. See methods and SI5 for details about EPR spectrum modeling. **(C)** Band energy diagram for both Oxygen terminated and SiO$_2$-coated diamonds with corresponding affinities ($\chi$) and bend alignments. The model shows the corresponding spatial dependence of the Fermi level ($E_f$), bottom of conduction band ($E_c$), top of the valence band ($E_v$), Nitrogen (P1), and NV center (NV$^{-/0}$). The energies and occupation of X-spins levels are also indicated **(D)** Schematic depiction of a bare (top) and a core-shell (bottom) structured diamond nanocrystal with an NV-qubit (orange) in the diamond host blue. Surface spins are indicated in yellow.

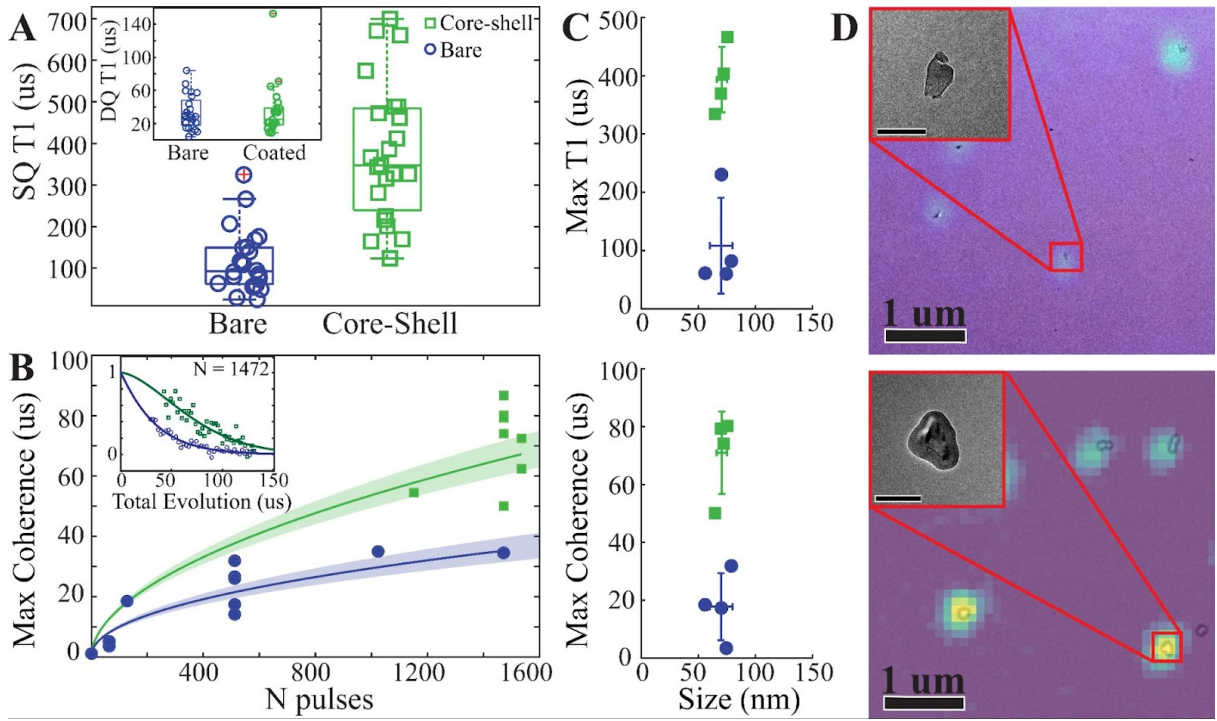

**Figure 3 — Relaxation and coherence of individually resolvable diamond nanocrystals. (A)** Single quantum relaxation measurements of bare (blue; $n = 24$) and core-shell (green; $n = 23$) particles. The inset shows double quantum relaxation times of the same particles (double quantum relaxations were measured at ~6.7G). **(B)** Maximally obtained $T_2$ times under CPMG dynamical decoupling of randomly selected twelve bare and eight core-shell particles as a function of the number of pulses applied. Note, for some of the investigated bare diamonds $T_2$ does not increase with $N$ (points with low $N$), while for all core-shell structured particles, we observed a $T_2$ for $N > 1,000$ (Fig. E8 and SI9). Solid lines are fits to a power law as described in the main text. The inset shows a representative coherence time trace data for one bare and core-shell particle with $N = 1472$. **(C)** Correlation of core size with $T_1$ (upper panel) and $T_2$ (lower panel) for four bare and four core-shell particles. **(D)** Representative CLEM images of bare (upper panel) and core-shell (lower panel) particles for the data shown in panel C. Insets are larger magnifications of individual particles (scale bar = 100nm).

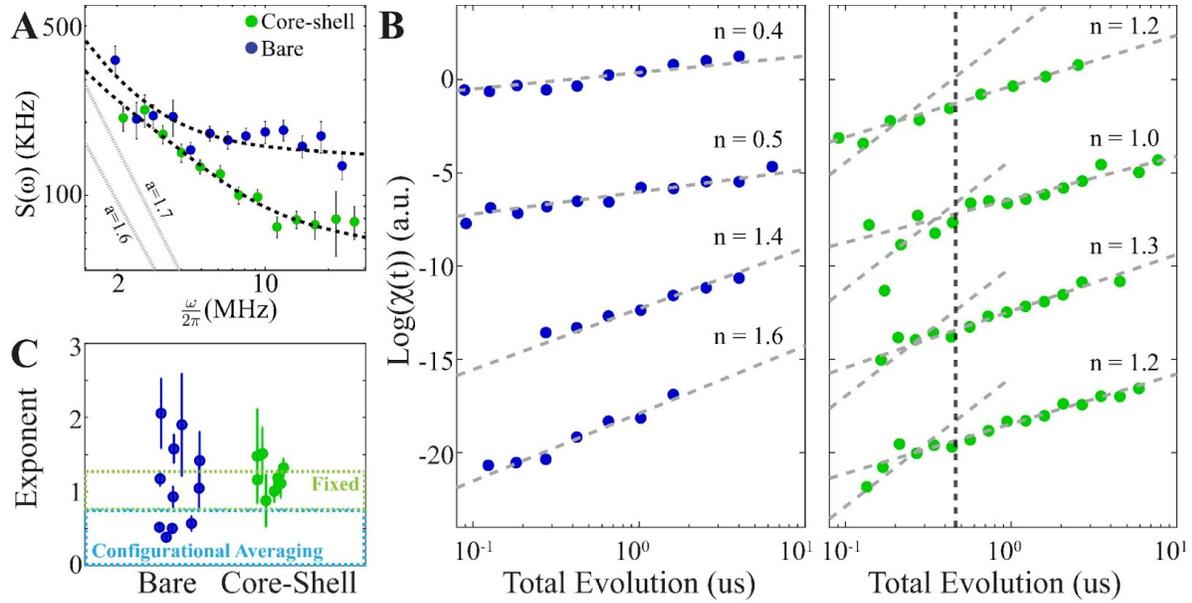

**Figure 4 — Probing spin bath properties using spectral decomposition and stretching factor spectroscopy. (A)** Spectral decomposition of CPMG data from bare (blue) and core-shell (green) particles. Gray dotted lines show fits to $1/f^a$ considering DQ data (see Fig. E9). Black dashed lines show fits for $1/f^a$ plus a single (for bare) or a double (for core-shell) Lorentzian, with the addition of white noise (see Fig. E9 for exact fitting details). **(B)** Four representatives bare (left panel) and core-shell (right panel) echo stretching factors. Gray dashed lines are exponential fits for the random walk regime (the cutoff is marked by the black dashed line). Core-shell particles also show ballistic regime fits to $n = 3$. **(C)** Distribution of echo stretching factors for bare (blue) and core-shell (green). Data points with $n \leq 0.75$ (cyan box) can be explained by configurational averaging, while data points with $n \approx 1$ (green box) correspond to fixed spins' positions (see SI11 for more details).

## Methods

### Diamond nanocrystals

40-45 nm diamond nanocrystals were obtained from Adámas Nanotechnologies Inc. In brief, type 1b microcrystals are manufactured by static high-pressure, high-temperature (HPHT) synthesis and contain about 100-200 ppm of substitutional N. These particles are milled, irradiated with 2-3MeV electrons, and annealed at 850 °C for 2 hrs by Adámas Nanotechnologies Inc[23].

### Synthesis of core-shell particles

The growth of Silica shells on diamond nanocrystals was performed using a sol-gel Stöber process[51] from a tetraethyl orthosilicate (TEOS) precursor. To ensure uniform shell growth and prevent aggregation we modified a Polyvinylpyrrolidone (PVP) based technique[52,24] (Fig. E1-2 and SI1).

### Characterization of particles

*Transmission Electron Microscopy.*

Bare and core-shell diamond nanocrystals were deposited on a Copper (formvar carbon film) or Silicon ($Si_3N_4$ film) grid while ensuring minimal aggregation (Fig. E3 and SI2). Images were taken using an FEI Tecnai G2 F30 300kV TEM.

*Size analysis.*

Particle sizes were analyzed with the help of ImageJ[53]. A 2-pixel Gaussian blur was applied before background subtraction. Thresholding was used to convert to a binary image from which the area ($A$) of an individually resolved crystal is calculated and subsequently converted to an equivalent diameter[54] (Fig. E3 and SI3). We analyze core-shell particles by using a two step thresholding process to separate the darker core from the brighter shell.

**Correlated light and electron microscopy (CLEM) measurements**

For CLEM measurements, nanocrystals were deposited on a silicon nitride TEM grid that was placed face-down on a glass coverslip with a fabricated coplanar waveguide. The sample was then placed in our home-built confocal microscope (SI4) for PL and coherence measurements. Subsequent TEM images enabled us to identify individual particles by comparing particle constellations in TEM with those obtained from confocal scans. As fiducial markers for alignment of the confocal and TEM images, we used the corners and edges of the TEM grid windows, followed by overlapping of bright PL spots to the nanocrystals' TEM pattern for fine alignment (see Fig. E3 for more details and CLEM images for PL and coherence measurements). Once we identified the particles of interest, high-magnification images were taken.

**Photoluminescence**

Photoluminescence (PL) was measured for bare and core-shell particles using our home-built confocal setup (SI4). By drop casting a mixture of bare and core-shell particles on the same TEM grid, we were able to ensure that both particle types were measured under identical conditions (Fig. E3). Correlation with TEM then enabled us to unambiguously identify each fluorescence spot as either a bare or a core-shell particle. A total of 93 individual particles (58 core-shell and 35 bare) were analyzed. The particle radius (see size analysis in methods and SI3) in Figure 1C was fitted to $ar^3$, where $r$ is the particle radius, and $a$ is a fit parameter. We extract $a = 0.65 \pm 0.11$ for bare, and $a = 1.20 \pm 0.16$ for core-shell particles.

**Spectrum analysis**

The photoluminescence (PL) spectrum of an ensemble of bare and core-shell particles with a core diameter of 70 nm was measured using Ocean Optics HR2000+ integrated into our home-built confocal microscope. We correct for variations in the particle density by normalizing the observed PL fluorescence spectrum ($PL(\lambda)$), i.e., $pl(\lambda) = PL(\lambda)/\Sigma_\lambda PL(\lambda)$. The normalized spectrum was then fitted to $pl = a\,pl^{(0)} + (1-a)\,pl^{(-)}$, where $a$ is a fit parameter with the restrictions $0 \leq a \leq 1$, $pl^{(0)}$ is the PL spectrum of NV$^0$ and $pl^{(-)}$ is that of NV$^-$.

**EPR measurements**

EPR measurements were performed using an X-band continuous wave EPR spectrometer from Bruker (Elexsys 500) with a 100-kHz field modulation and a high-quality resonator ER 4122 SHQE. EPR spectra were collected with an incident microwave power of 2 mW, and a modulation amplitude of 0.2 mT, at room temperature. Experimental EPR spectra were decomposed into individual resonances. Fitting was done with the help of the EasySpin55 software package which allowed us to determine spectroscopic parameters for each individual EPR signal. Measurements were done on 70 nm particles to ensure that the average particle matches the ones that were measured for coherence. Results for bare and core-shell particles were normalized both by mass and number of particles so signals for both samples could be properly compared. See Fig. E4 and SI5 for details about the normalization procedure as well as results for 40nm particles.

**Electronic band structure**

The band structures for diamond nanocrystals and amorphous Silica were derived from literature values. We considered a type 1b (~100 ppm N impurities) oxygen-terminated diamond nanocrystal with a positive electron affinity of ~2eV and a band gap of 5.5eV[34]. The bulk conduction band, $E_C$, lies approximately 1.5eV below the vacuum level[33,56] resulting in a ~0.5eV downward band bending[57]. The formation energy of P1 centers is positioned 1.7eV below $E_C$ and only 0.1eV below the Fermi level[33,56,58]. The NV⁻ band is located 2eV above the valence band. The band structure for amorphous Silica featured a large, 9.4eV band-gap, $E_C$ that is located 0.7eV below the vacuum level, and a fermi level that lies 5.55eV above the valence band[59,60]. Equalization of the fermi level ($\Delta E_f = 1.45 eV$) produced the band bending for alignment at the heterojunction. See Fig. E5 and SI6 for more details.

*Simulation*

The band bending for bare and core-shell diamond nanocrystals is obtained by solving the Poisson's equation that reflects the charge density arising from electrons, holes, P1 (or ionized Nitrogen), and charged states of both vacancies and NV centers. The boundary condition for the electrostatic potential is obtained by assuming charge neutrality deep in the crystal, the formation energies and densities of defects are obtained from literature[23,61–63], and the band bending values at the surface as extracted from the model in Fig. 2C. For more details on the simulation, see Fig. E6 and SI6.

**NV-based double electron-electron resonance (DEER) measurements**

To confirm the coupling of these paramagnetic species to the NVs in our particles, we performed NV-based DEER experiments following the protocol described in refrence[18], Fig E7, and SI7. By normalizing the contrast of a DEER measurement with respect to Hahn echo decay we remove contributions from other noise sources. Rabi frequency of the paramagnetic defects, needed for DEER free induction decay (FID), was measured using a correlation-based sequence described in[64]. All measurements were performed at ~205G.

**Coherence and relaxation measurements**

All measurements were done in 2 different batches to ensure reproducibility. The first batch was measured using nanocrystals deposited on a #1.5 glass coverslip. The second batch was measured using nanocrystals deposited on silicon nitride TEM window grids (PELCO 15nm or 50nm Si3N4, for CLEM). Nanocrystals for measurement were chosen without favoring brighter emitters in the confocal imaging. We found this is crucial in order to ensure a representation of single nanocrystals with similar core sizes, as confirmed by CLEM (Fig. 3D and Fig E3).

*T₁ measurements*

SQ and DQ relaxation measurements were performed using a sequence adopted from Myers et al.[40]. Setting $\gamma$ as the DQ transition rate and $\Omega$ as the SQ transition rate, we can extract $T_1^{SQ} = \frac{1}{3\Omega}$ and $T_1^{DQ} = \frac{1}{\Omega+2\gamma}$ (see SI8 for details). All relaxation measurements were done at low magnetic field (~6.7G) using 24 bare and 23 core-shell particles.

*T₂ measurements*

Coherence measurements were performed using Hahn-Echo and CPMG sequences illustrated in Fig. E8A (inset). Randomly selected bare ($n = 12$) and core-shell ($n = 8$) diamond nanocrystals from two different batches (glass and silicon nitride grids) were measured for SQ and DQ relaxation times at low fields. The external magnetic field was then aligned to ~185G for $T_2$ measurements starting with $N = 1$ (echo). $T_2$ was measured with an increasing number of dynamical decoupling pulses up to $N > 1000$ or saturation (SI9). The max $T_2$ times and corresponding $N$ pulses are plotted in Fig. 3B. Figure E8 shows the CPMG data collected from all the particles for $T_2$ echo and $T_2$ max.

*$T_2$ fitting*

Coherence data, $C(t)$, was fitted to a stretched exponential decay of the form $C(t) = ae^{-\chi(t)}$, where $\chi(t) = (t/T_2)^n$, and $a$, $T_2$, and $n$ are fit parameters. To account for pulse evolution time, we set the initial time, $t_0$, to be equal to the total time of π pulses, such that $t_0 = Nt_\pi$ (see Fig. E8B and SI9 for a numerical simulation validating this protocol). To account for pulse errors, we force the fitting parameter $a$ to be monotonous decreasing in $N$. It is important to note that we are using $n$ as a fitting parameter and not based on a model. Fixing $n = 3$, as would be the case of a ballistic phase evolution in a fixed spin configuration under dynamical decoupling, would significantly overestimate $T_2$.

**Noise spectral density**

*Theory*

The coherence of a qubit is described by $C(t) = exp[-\chi(t)]$, where $t$ is the total free precession time. Under CMPG decoupling with $N$ π-pulses, $\chi(t)$ is given by:

$$\chi(t) = -\frac{1}{2}\int_0^\infty \frac{d\omega}{2\pi} S(\omega)|\lambda(\omega,t)|^2,$$ where $|\lambda(\omega,t)|^2$ is the filtering function[65,66]. For $N \gg 1$, the filtering function presents a primary peak at $\omega = \omega_0 = N\pi/t$, yielding $C(t) \approx exp[-S(\omega_0)t]$ (SI11). Accordingly, measuring the coherence data for different $N$ allows us to extract the spectral noise density at different frequencies, $\omega_0$. This method is applied for both bare and coated diamonds, with corresponding $S(\omega)$ shown in Fig. 4A. We emphasize, however, that this method assumes linear dependence of $\chi(t)$ with respect to $t$.

*Estimating $\chi(t)$ from experimental results*

To ensure the condition $N \gg 1$ applies, we limit our analysis to CPMG experiments with $N > 64$. Experimentally obtained CPMG time traces (Fig. 3B inset and Fig E8) were normalized and data points larger than 1 and smaller than 0 were discarded. The spectral density was extracted from these data using the approximation $\chi(t) = tS(\omega)/\pi$. The spectrum was binned into 14 logarithmic bins and plotted in Fig. 4A. See Fig. E9 for data before binning.

*Plotting DQ, CPMG, and SQ on the same scale*

The DQ and SQ relaxation measurements (CPMG dephasing and SQ relaxation) are sensitive to different noise sources, i.e., DQ relaxation is sensitive to transverse electric fields at frequencies of $18.8 MHz$ whereas SQ relaxation is sensitive to transverse magnetic at $2.87 GHz$ and parallel electric fields . To compare QD and SQ elaxation measurements with our noise spectroscopy we follow the procedure described in reference[40]. See SI10 for more details.

*Fitting*

Using our measured CPMG and DQ data we can now fit a model for the noise spectrum. We follow a modified version of the procedure described in reference[40].

The noise spectrum of a spin bath is expected to follow a Lorentzian with the generic form:

$$S(\omega) = \sum_k \left( \frac{\Delta_k^2 \tau_{C,k}}{\pi(1+(\omega \tau_{C,k})^2)} \right), \qquad \text{(eq. 1)}$$

where $\Delta_k$ and $\tau_{c,k}$ are the coupling strength and bath correlation time, respectively. The power spectrum corresponding to electric noise is expected to be described by $1/f^a$-noise[4,18,40,67,68]. The full fitting function for the combined noise is then given by:

$$S(\omega) = \sum_k \left( \frac{\Delta_k^2 \tau_{C,k}}{\pi(1+(\omega \tau_{C,k})^2)} \right) + \frac{\Delta_e}{\omega^a}. \qquad \text{(eq. 2)}$$

$S_{DQ}(\omega) = \gamma(\omega)$ was obtained as discussed in the *$T_1$ measurements and fitting* section of the methods. The two other parameters, $\Delta_e$ and $a$, are related to each other, such that, $\Delta = S_{DQ}(\omega_{DQ}) * \omega_{DQ}^a$, where $a$ is a fit parameter. The fitting process and results are illustrated in Figure E9.

**Echo exponent analysis**

A detailed derivation of the functional form of the exponent of the coherence factor can be found in SI11. The obtained stretching factor is the result of Ising interaction between the NV and a D-dimensional fluctuating spin bath[39,42,43,67,69].

In the following we consider two separate scenarios:

First, we consider a Lorentzian spin bath where the spin position remains fixed in space. The stretching factor then takes the functional dependency:

$$\chi(t) \propto \begin{cases} t^3 & t \ll \tau_C \\ t & t \gg \tau_C \end{cases}.$$

Second, we consider a Lorentzian spin bath where the spin position does not remain fixed and moved over time. Such a scenario is expected if paramagnetic centers of the bath can be ionized under laser excitation. The resulting stretching factor is then given by:

$$\chi(t) \propto \begin{cases} t^{3D/2\alpha} & t \ll \tau_C \\ t^{D/2\alpha} & t \gg \tau_C \end{cases},$$

where $D$ is the spatial dimensionality of the fluctuators and $\alpha$ is the interaction's scaling with distance ($\alpha = 3$ for dipole interactions and $2$ for point-like charge interactions). Note, in our analysis we consider both scenarios.

*Fitting and plotting χ(t) time trace plots*

A transition from the ballistic to the random walk regime occurs at $\tau_c$, and leads to a double exponential decay for the echo signals. However, in a system with short correlation times, as it is the case in our diamond nanocrystals (see Fig. 4A and Fig. E9), we would expect to be primarily in the random walk regime. This prevents us from using our data for an accurate fitting in the ballistic regime. To ensure analysis within the random walk regime we only consider evolution times larger than $10\tau_c$, which allows us to fit this truncated data to a single exponential. The bare and core-shell echo time traces were plotted (Fig. 4C and Fig. E10) using the identity $Log(\chi(t)) = Log(-Log(C(t)))$. For core-shell particles in the ballistic regime we also plotted $\chi(t) = (\frac{t}{T_2})^3$ as a guide to the eye.

**Statistics**

All error bars represent standard error unless otherwise noted. $T_1$ relaxation plot in Fig. 3A is illustrated using a box and whisker diagram showing the minimum, lower quarter (25th percentile), median, upper quarter (75th percentile), and maximum points of the given data. Outliers are marked with a red + and correspond to points that are more than $q_3 + 1.5(q_3 - q_1)$ or less than $q_1 - 1.5(q_3 - q_1)$ where $q_1$ and $q_3$ are the 25th and 75th percentiles of the sample data, respectively. A t-test for the bare ($n = 24$) and core shell ($n = 23$) $T_1$ data sets (Fig. 3A) resulted in $p = 8.5 \times 10^{-8}$. A t-test for the bare ($n = 11$) and core shell ($n = 8$) $T_2$-echo data sets (Fig. E8 and Fig. E10) resulted in $p = 0.03$. A t-test for the bare ($n = 12$) and core shell ($n = 8$) $T_2$ max data sets (Fig. 3B) resulted in $p = 4.6 \times 10^{-7}$. All analyses were done using two tails unequal variance T-test.

## Data availability

The data that support the finding of this study are available from the corresponding authors upon reasonable request.

## Methods references

**Acknowledgments**

We thank Ania Jayich and Shimon Kolkowitz for insights related to surface charge noise; Nathalie de Leon, Nazar Delegan, Joseph Heremans, Olga Shenderova, and Alexander Zaitsev for discussions on surface terminations; Mouzhe Xie for help on XPS; and Susumu Takahashi and Karoly Kolczer for discussions on EPR. U.Z., and P.C.M. acknowledge financial support from the NSF Grant No. OMA-1936118 and NSF QuBBE QLCI (NSF OMA- 2121044). DEER measurements were performed by X.Y. and S.W. with support from US DOE, Basic Energy Sciences, Division of Chemical Sciences, Geosciences, and Biosciences, though ANL under Contract No. DE-AC02-06CH11357. D.R.C. and M.E.F. were supported as part of the Center for Molecular Quantum Transduction (CMQT), an Energy Frontier Research Center funded by the U.S. Department of Energy, Office of Science, Basic Energy Sciences under Award No. DE-SC0021314. I.G. and D.V.T. acknowledge support by the National Science Foundation under Grant DMR-2019444 (IMOD an NSF-STC). We acknowledge the use of the Pritzker Nanofabrication Facility at the University of Chicago (NSF ECCS-2025633), the University of Chicago Materials Research Science and Engineering Center (DMR-2011854), as well as the assistance of Yimei Chen, Dr. Jotham Austin, and Dr. Tera Lavoie from The University of Chicago Advanced Electron Microscopy Core Facility (RRID:SCR_019198). EPR measurements were performed at the Center for Nanoscale Materials, a U.S. Department of Energy Office of Science User Facility, supported by the U.S. DOE, Office of Basic Energy Sciences, under Contract No. DE-AC02-06CH11357. The XPS and Raman spectroscopy made use of the EPIC, Keck-II, and/or SPID facility(ies) of Northwestern University's NUANCE Center, which has received support from the Soft and Hybrid Nanotechnology Experimental (SHyNE) Resource (NSF ECCS-1542205); the MRSEC program (NSF DMR-1121262) at the Materials

## Authors contributions

U.Z. and A.R.J. designed and built the experiment. U.Z., A.R.J., and L.C. performed the sample preparation, coherence, and TEM measurements and analyzed the data. D.R.C. and M.E.F. developed the theoretical model and D.R.C. performed the theoretical calculations. A.W. helped with material synthesis. I.G. performed EPR and X.Y. and S.W. performed DEER measurements. D.V.T., M.E.F., A.E.K., and P.C.M. provided guidance. P.C.M. developed the idea and supervised the project. All authors contributed to discussing the results and writing the paper.

## Competing interests

The authors have filed a provisional patent application for work described in this manuscript.

## Additional information

Supplementary information is available for this paper.

Correspondence and requests for materials should be addressed to Peter C. Maurer.

**Extended Data**

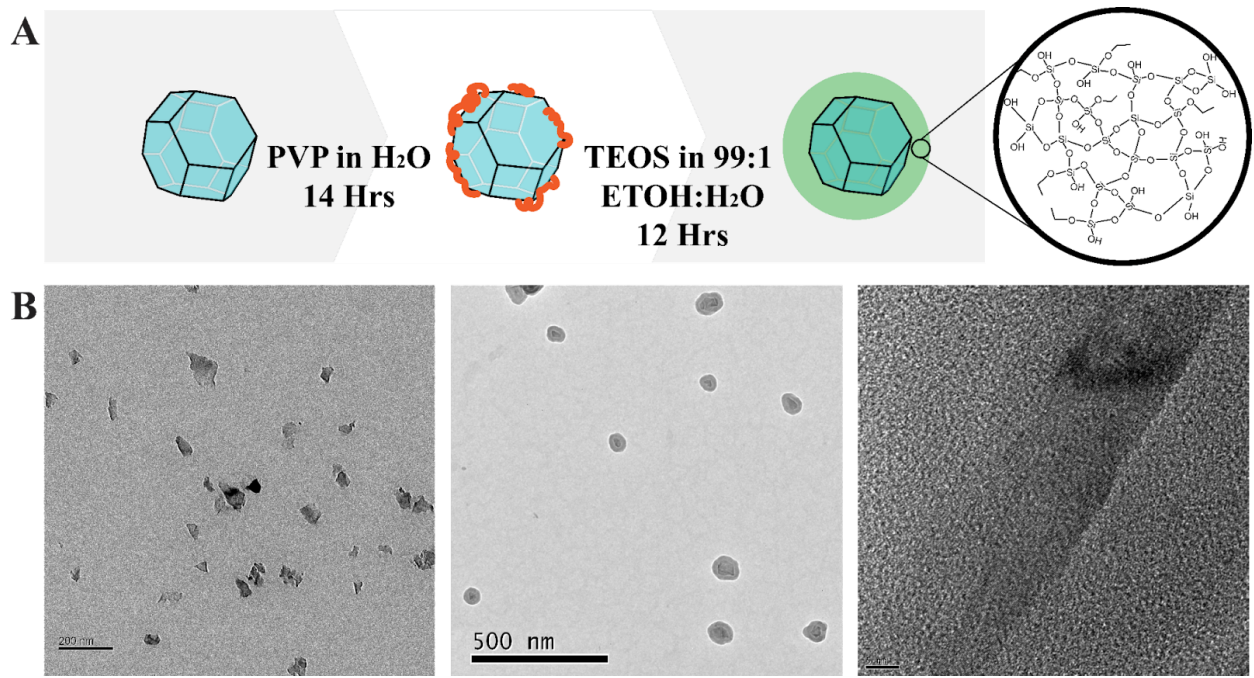

**Figure E1 — Growth of silica shell on diamond nanocrystals (A)** Two-step shell growth process including a PVP stabilization step and a sol-gel growth step. The enlarged circle to the right illustrates the expected amorphous silica structure containing alcohols and ethyl groups. **(B)** TEM images of bare (left panel) and core-shell (middle panel) particles. A high-magnification image (right panel) of a diamond nanocrystal encapsulated in silica featuring the diamond crystal lattice lines. Detailed protocol for shell growth can be found in SI1.

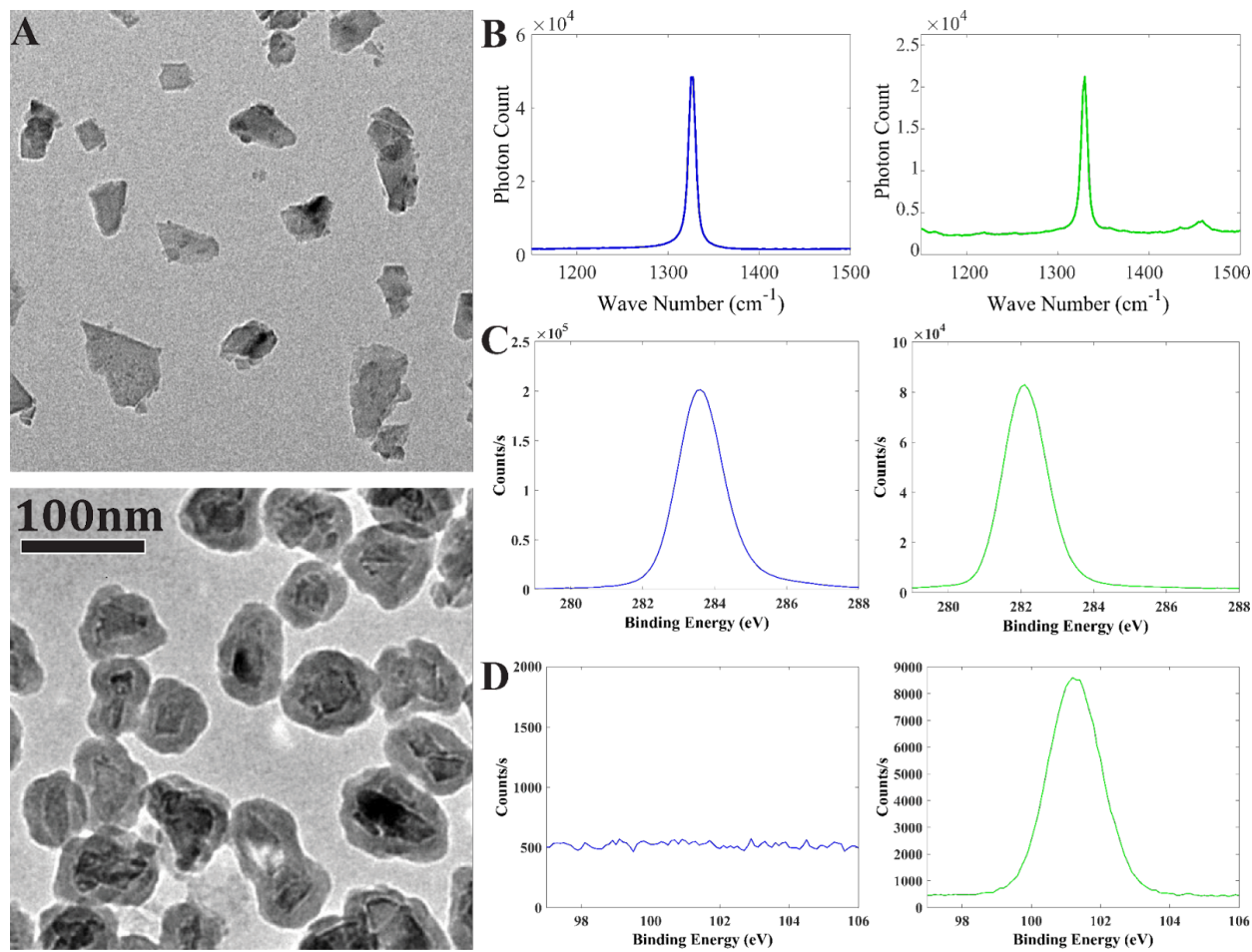

**Figure E2 — Characterization of bare and core-shell diamond nanocrystals. (A)** Original TEM images for Fig. 1A. **(B)** Raman spectrum of bare (left panel) and core-shell (right panel) particles featuring diamond peak at 1332cm$^{-1}$. **(C)** XPS spectrum of bare (left panel) and core-shell (right panel) particles featuring diamond peak at ~284eV. **(D)** XPS spectrum of bare (left panel) and core-shell (right panel) particles featuring silica peak at ~101eV.

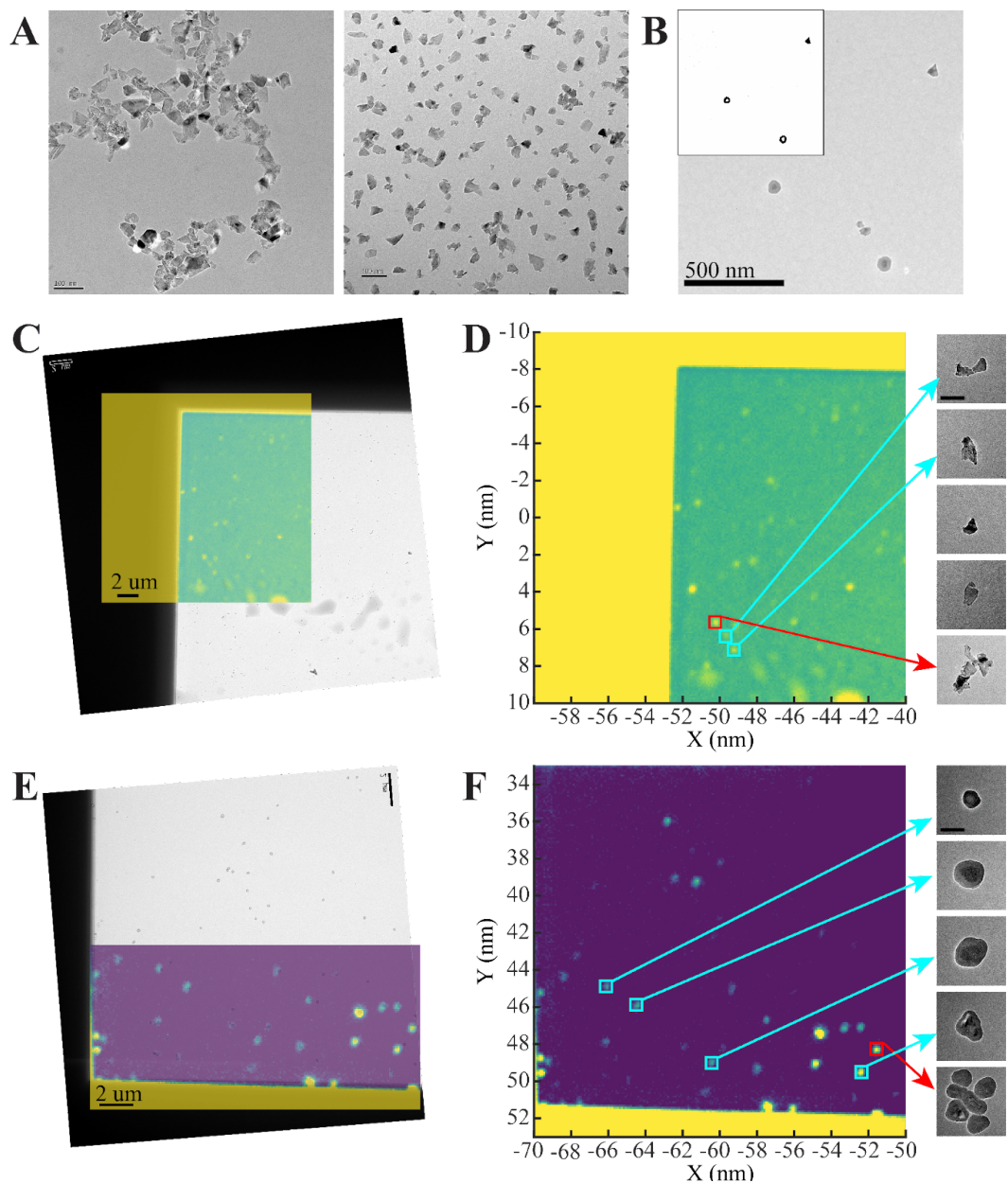

**Figure E3 — TEM grid processing for CLEM and software size analysis. (A)** TEM image of -bare particles deposited on a UV-ozone treated grid (left panel) and a UV-ozone + PEI treated grid (right panel). **(B)** TEM image of bare and core-shell particles on the same grid. Inset shows the binary image produced after processing and thresholding for area calculations. Only fluorescent particles were processed. **(C)** Alignment of TEM and confocal image for CLEM of bare particles used in Figure 3. **(D)** Enlarged confocal image containing 2 bare particles that were used in Figure 3 (cyan) and 1

measurement that was disqualified after classification as an aggregate (red). Note that the other 2 particles used in Figure 3C are found on a separate confocal scan featured in Figure 3D. TEM images of individual particles are shown in the panels to the right. **(E)** Alignment of TEM and confocal image for CLEM of core-shell particles used in Figure 3. **(F)** Enlarged confocal image containing all the core-shell particles that were used in Figure 3 (cyan), including 1 measurement that was disqualified after classification as an aggregate (red). TEM images of individual particles are shown in the panels to the right.

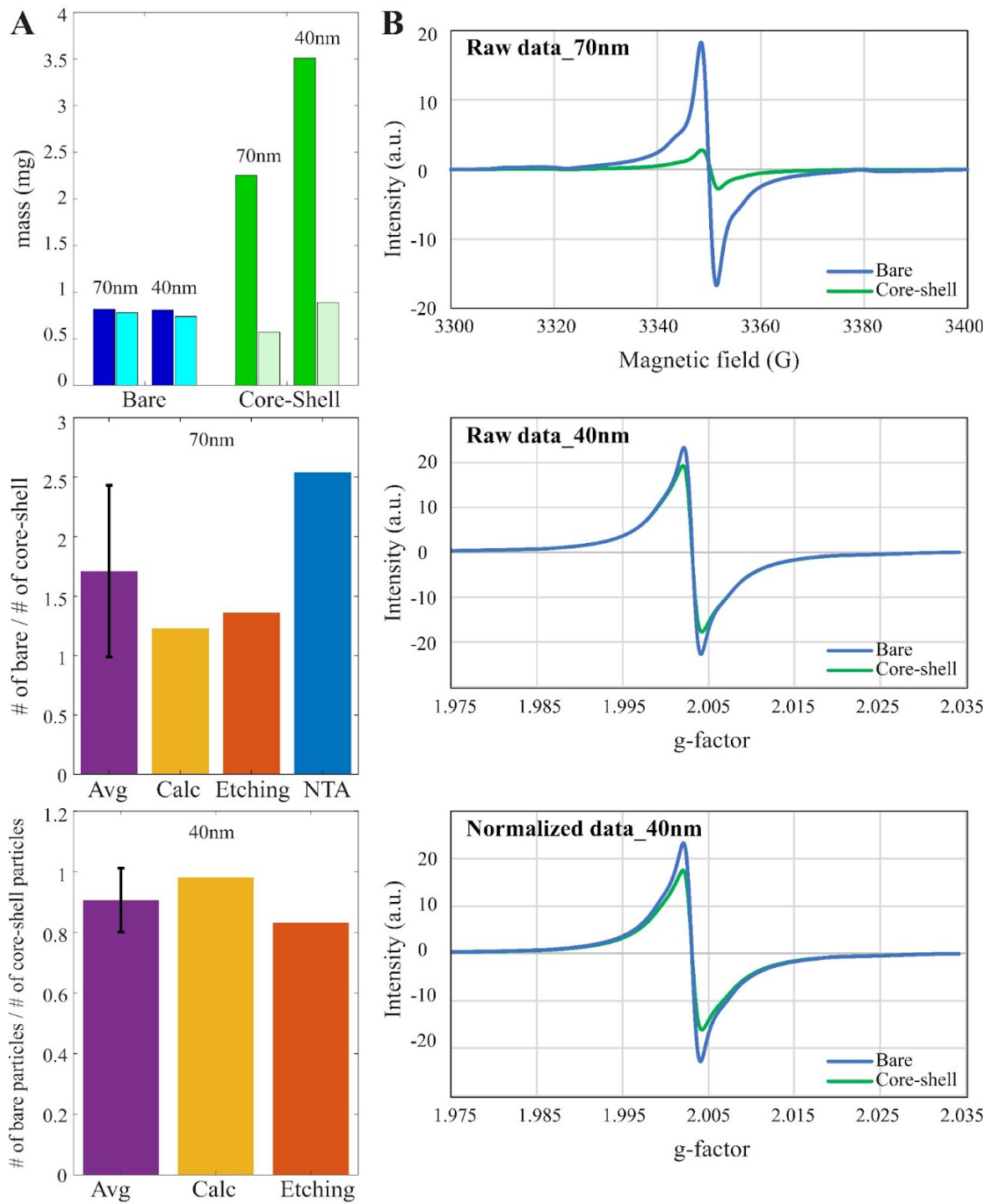

**Figure E4 — EPR of bare and core-shell particles. (A)** Normalization of EPR signals. The upper panel shows the mass of bare (blue) and core-shell (green) 70 and 40nm particles before (dark shades) and after (light shades) etching with KOH. The lower two panels show all normalization factors, $F = N_{bare}/N_{CS}$ for 70nm (middle panel) and 40nm (lower panel) particles (see SI5 for a detailed description of the normalization procedure). **(B)** Additional EPR data for bare (blue) and core-shell (green) particles. The upper panel shows raw EPR signals for 70nm particles before normalization (see Fig. 2A, B for the normalized signal). The lower two panels show the raw EPR for 40nm particles (middle panel) and the normalized signal (lower panel). Resonant frequencies for 70 nm particles were 9.392094GHz and 9.392505GHz for bare and core-shell particles, respectively. Resonant frequencies for 40 nm particles were 9.395766GHz and 9.392509GHz for bare and core-shell particles, respectively.

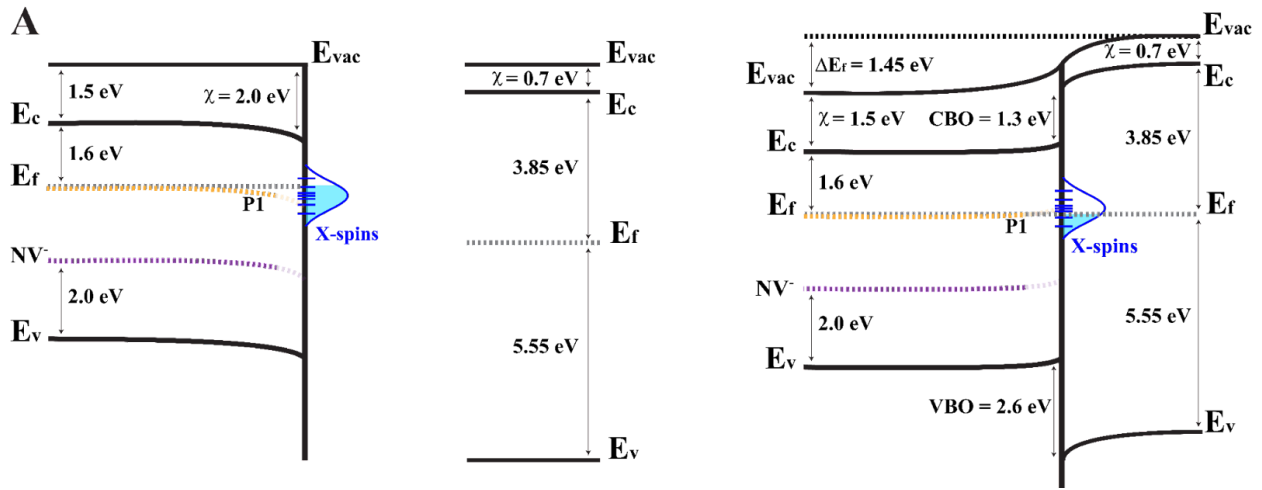

**Figure E5 — Electronic band structures.** The electronic band structure of diamond nanocrystals (left), amorphous silica (middle), and the heterojunction of the core-shell particles (right), leading to the depletion of paramagnetic P1 and X-spins after encapsulation in a silica shell. The modeled bending at the diamond interface is $-0.5 eV$ and $0.225 eV$ for bare and core-shell particles, respectively. Note that while the illustrated bending in Figure 2C is exaggerated for better visibility, here it is illustrated true to scale.

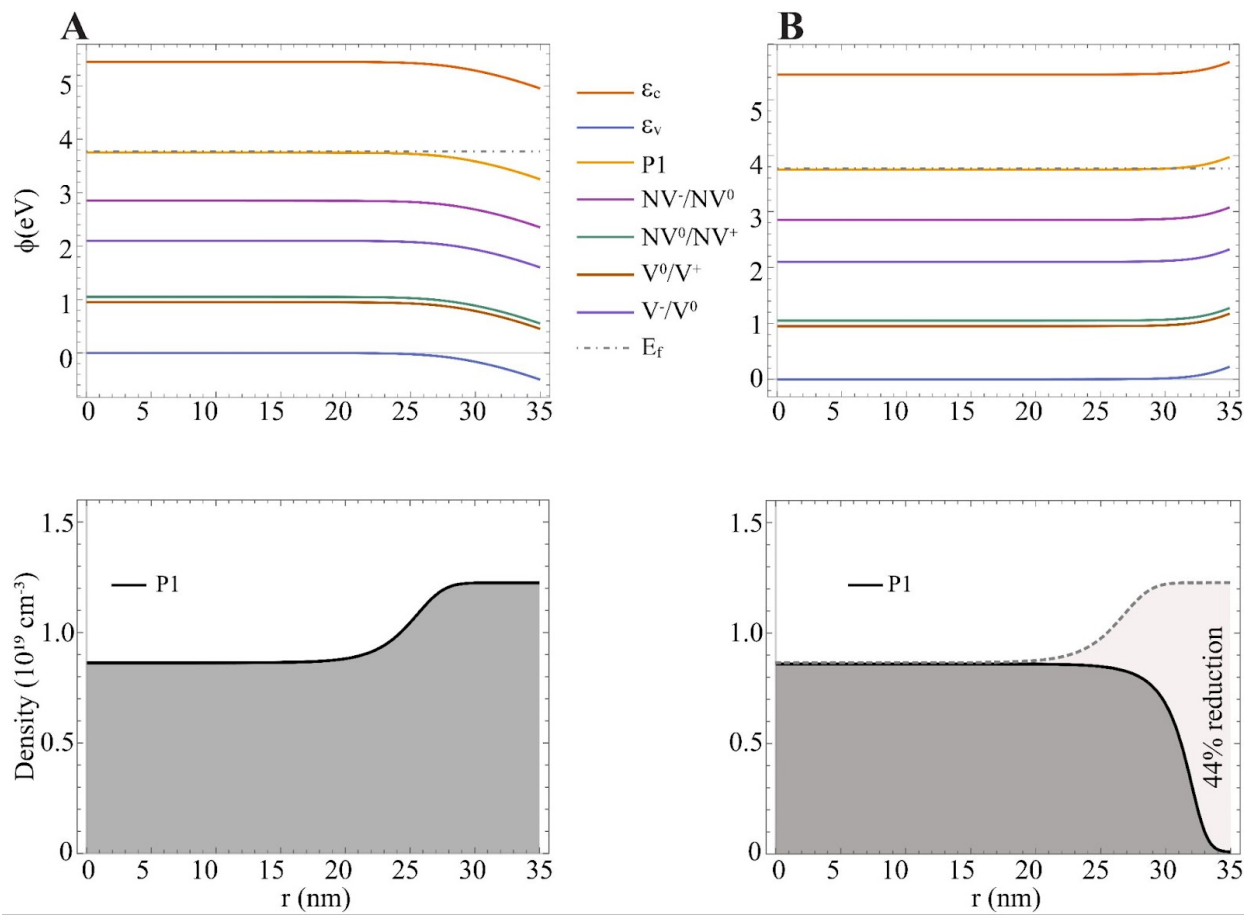

**Figure E6 — Band bending simulation. (A)** Simulation results showing the band diagram (upper panel) and the volumetric density of paramagnetic P1 centers (lower panel) as a function of distance from the center, $r$, in bare diamond nanocrystal with oxygen termination. **(B)** Simulation results showing the band diagram (upper panel) and the volumetric density of paramagnetic P1 centers (lower panel) as a function of $r$ for core-shell nanocrystal. The expected change in P1 is shown in light gray demonstrating a 44% reduction. See SI6 for simulation details.

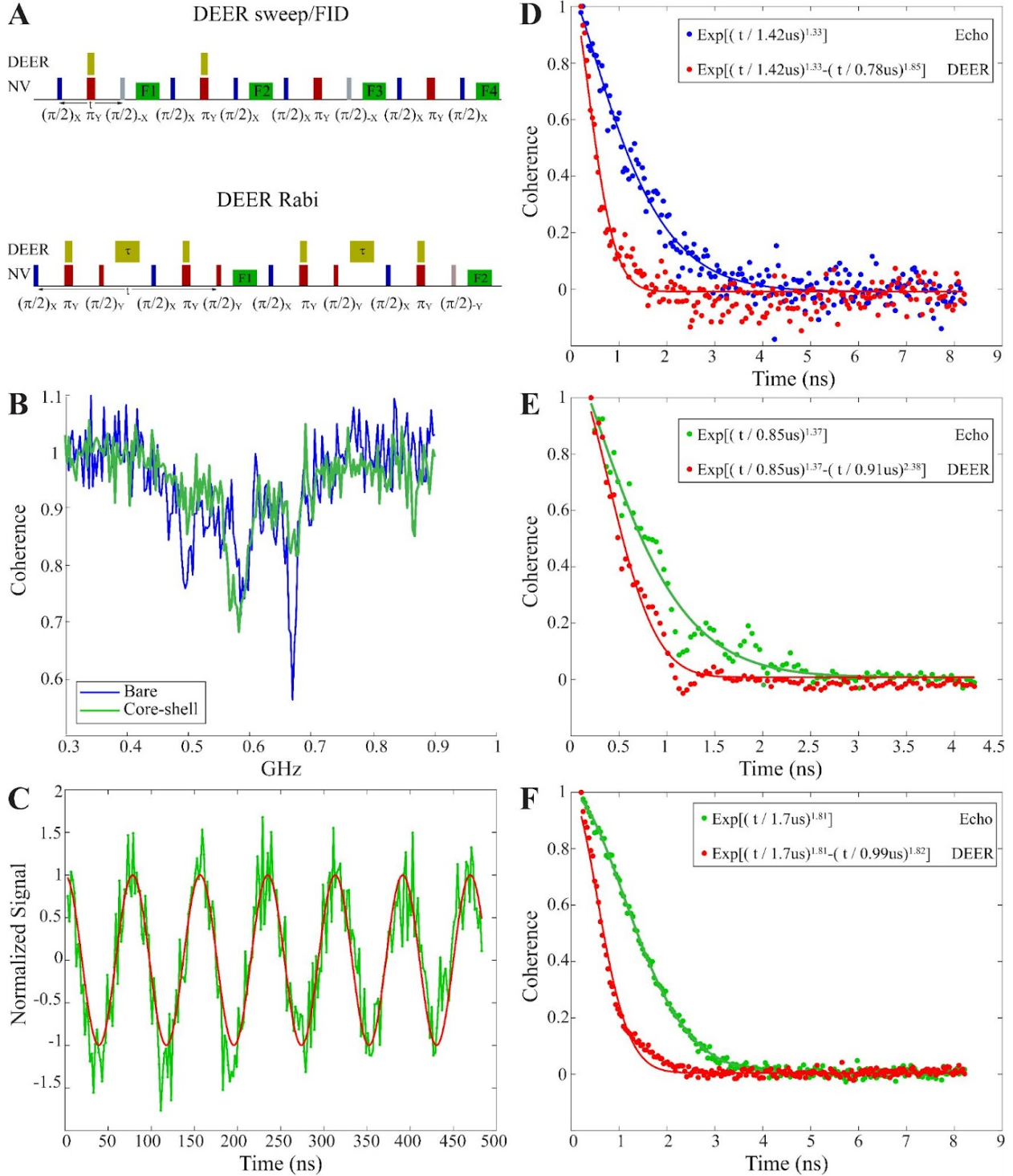

Figure E7 — DEER measurements. (A) Sequences used for DEER resonance\FID (top), and DEER Rabi (bottom) measurements. The DEER resonance\FID sequence contains four parts. The first two parts are responsible for DEER decay and end with a 180° phase-flipped π/2 pulse to cancel common mode noise (e.g. fluorescence decay due to charge instability). The last two parts of the sequence are responsible for the NV Hahn echo. We use the DEER decay signal, $S_D = (F1 - F2)/(F1 + F2)$, and the Hahn-Echo decay signal, $S_E = (F3 - F4)/(F3 + F4)$ to extract the DEER FID signal defined as: $S_{FID} = S_D/S_E$. To find the DEER resonance, we fixed $t = 400 ns$, while sweeping the microwave frequency. To extract the FID time trace, we set the microwave frequency to the resonance of the paramagnetic defects, while sweeping $t$. The duration of the π pulse is determined by extracting the DEER Rabi frequency. In the Rabi measurement, we fix $t$, while sweeping the duration of the middle DEER pulse. The signal is defined as $S_{Rabi} = (F1 - F2)/(F1 + F2)$. The final NV π/2 pulse is phase flipped to cancel common mode noise. (B) Detected DEER resonances for bare (blue) and core-shell (green) particles at ~0.498GHz, ~0.591GHz, and ~0.669GHz, correspond to expected values for P1 centers and X-spins. (C) DEER Rabi in a core-shell particle confirms coherent driving of paramagnetic defects at 12.8MHz. (D) Spin echo (blue) and DEER FID (red) of a bare particle. Inset showing fit results with DEER FID of 0.78µs. (E-F) Spin echo (green) and DEER FID (red) of core-shell particles. Inset showing fit results with DEER FID of 0.99µs (E) and 0.91µs (F). The particle in panel (E) presents nuclear spin oscillations likely resulting in shorter spin echo time. All measurements were performed at ~205G. See SI7 for more details.

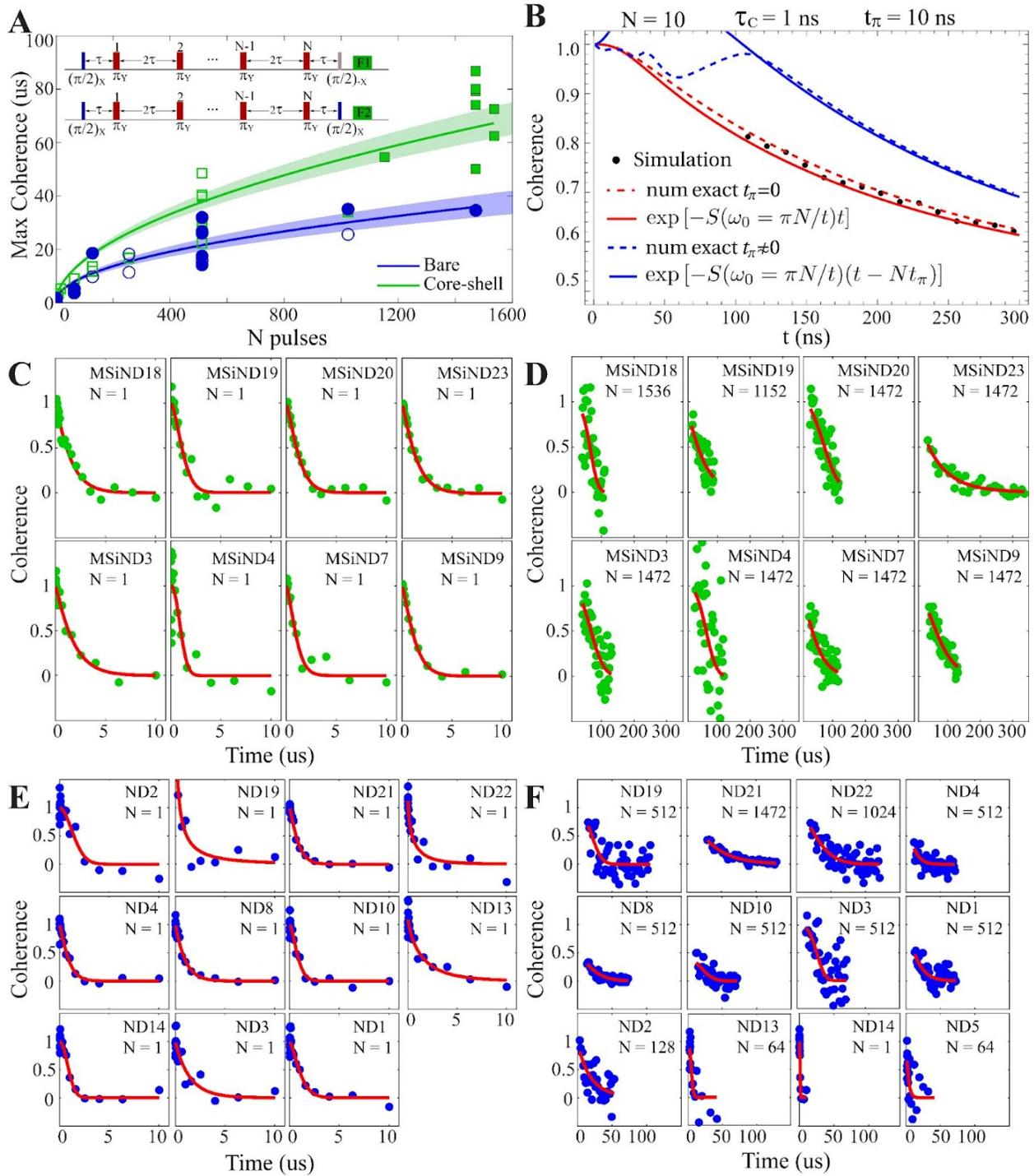

**Figure E8 — T2 measurements. (A)** A full version of Figure 3B featuring all CPMG measurements for bare (blue) and core-shell (green) particles. Fully colored markers represent the maximum T2 measured

for a given particle. Solid lines are fits to $T_2(N) = T_{2,echo} N^k$, considering all measurements for a given group. Since all core-shell particles showed measurable improvement for the highest number of pulses measured ($N>1000$), the fitted parameter $k = 0.53$ is a representation of the average core-shell particle. Only 2 bare particles continued to show improvement for $N>1000$, and most saturated at a much smaller number of pulses. Therefore, the fitted parameter $k = 0.47$ represents a higher bound for bare particles that showed a large distribution of $k = 0$ to $0.47$. The sequence used for CPMG measurements is shown in the inset (initialization pulse was ignored). The T2 decay signal was recorded as $S_{CPMG} = (F1 - F2)/(F1 + F2)$ and fitted to the time-dependent coherence, $C(t) = ae^{-\chi(t)}$, where we defined $\chi(t) = (\frac{t}{T_2})^n$. **(B)** Monte Carlo simulation assessing the effect of dephasing during the $\pi$-pulses. We set $N = 10$, $\tau_C = 1ns$, and $t_\pi = 10ns$ to represent the scenarios in our CPMG measurements where $\tau_C \leq t_\pi$ and $Nt_\pi$ is on the same order of magnitude as $T_2$ (see SI9 for more simulation details). The exact numerical (dashed curves) and the $C(t) \approx exp[-S(\omega_0)t]$ approximated (solid curves) coherence signals were calculated assuming no dephasing during the pulse duration (blue) and for an infinitesimally short pulse, $t_\pi = 0ns$ (red). **(C)** T2 echo signals for the eight measured core-shell particles, with an average $T_2^{Echo} = 1.42 \pm 0.20 \mu s$. **(D)** T2 CPMG signals for the eight measured core-shell particles. **(E)** T2 echo signals for eleven measured bare particles (ND5 Echo $T_2$ wasn't recorded due to an error), with an average $T_2^{Echo} = 1.05 \pm 0.41 \mu s$. **(F)** T2 CPMG signals for the twelve measured bare particles.

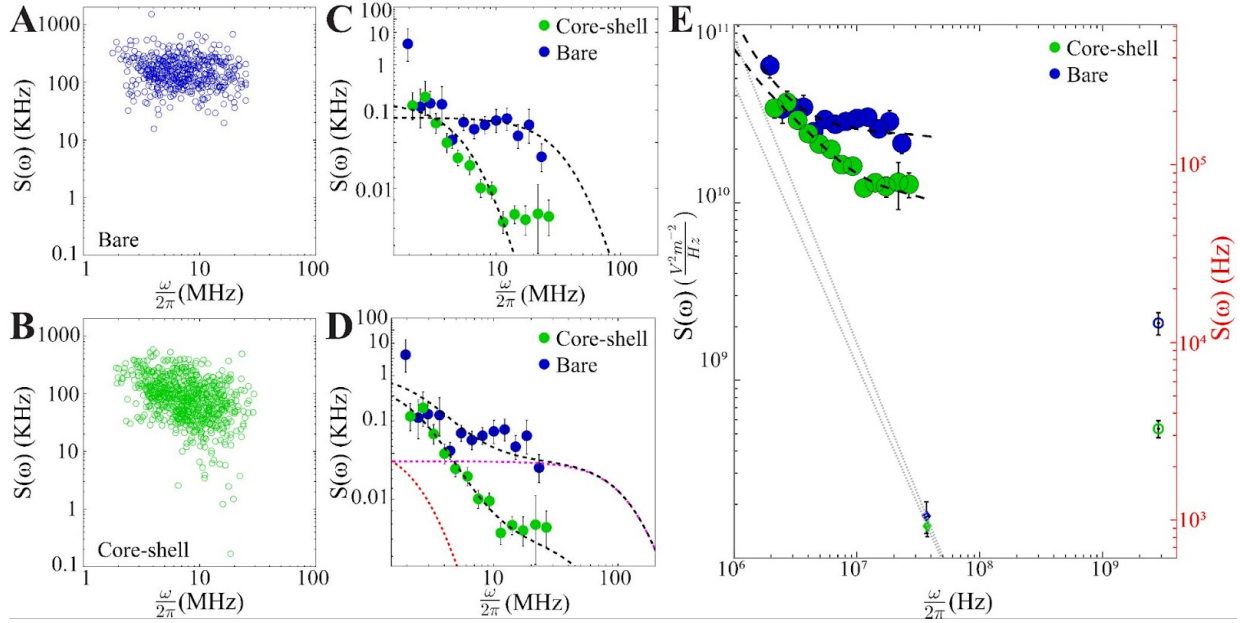

**Figure E9 — Noise spectrum fitting.** All points obtained with the noise spectral density extraction procedure for bare **(A)** and core-shell **(B)** particles before binning. **(C)** The data was grouped to 14 logarithmic bins. A single Lorentzian fit to equation (1) did not fit well for both groups (bare: $\Delta = 1.4e7$, $\tau_C = 3ns$, $rsquare = 0.06$; core-shell: $\Delta = 6.1e6$, $\tau_C = 17ns$, $rsquare = 0.63$). **(D)** A double Lorentzian fit described well only the core-shell group (bare: $\Delta_1 = 3.4e6$, $\tau_{C1} = 44ns$, $\Delta_2 = 2.2e7$, $\tau_{C2} = 0.9ns$, $rsquare = 0.53$; core-shell: $\Delta_1 = 3.8e6$, $\tau_{C1} = 46ns$, $\Delta_2 = 1.1e7$, $\tau_{C2} = 1.7ns$, $rsquare = 0.96$). Magenta and red dashed lines represent the bare particles' fit for the Lorentzians with short and long $\tau_C$, respectively. It is clear that the short $\tau_C$ Lorentzian is dominant, and that a good fit can only be obtained by adding a Lorentzian with much longer $\tau_C$, in the regime dominated by electric noise. **(E)** A fit for both groups was obtained by considering electric field noise and fitting $1/f$-like noise using the DQ relaxation data and equation (2). (bare: $\Delta = 2.4e7$, $\tau_C \leq 1ns$, $a = 1.7$, $rsquare = 0.75$; core-shell: $\Delta_{C,1} = 2.9e6$, $\tau_{C,1} = 40ns$, $\Delta_{C,2} = 1.3e7$, $\tau_{C,2} \leq 1ns$, $a = 1.6$, $rsquare = 0.95$). The fitted $a$ is within the expected range, $1 \leq a \leq 2$, arising from one or more Lorentzian noise baths, and agrees well with previous reports[40]. All dashed black lines are the fits to the total data. Only CPMG data for pulse number $N > 64$ was considered to uphold the approximation of the filtering function as a delta function (SI9).

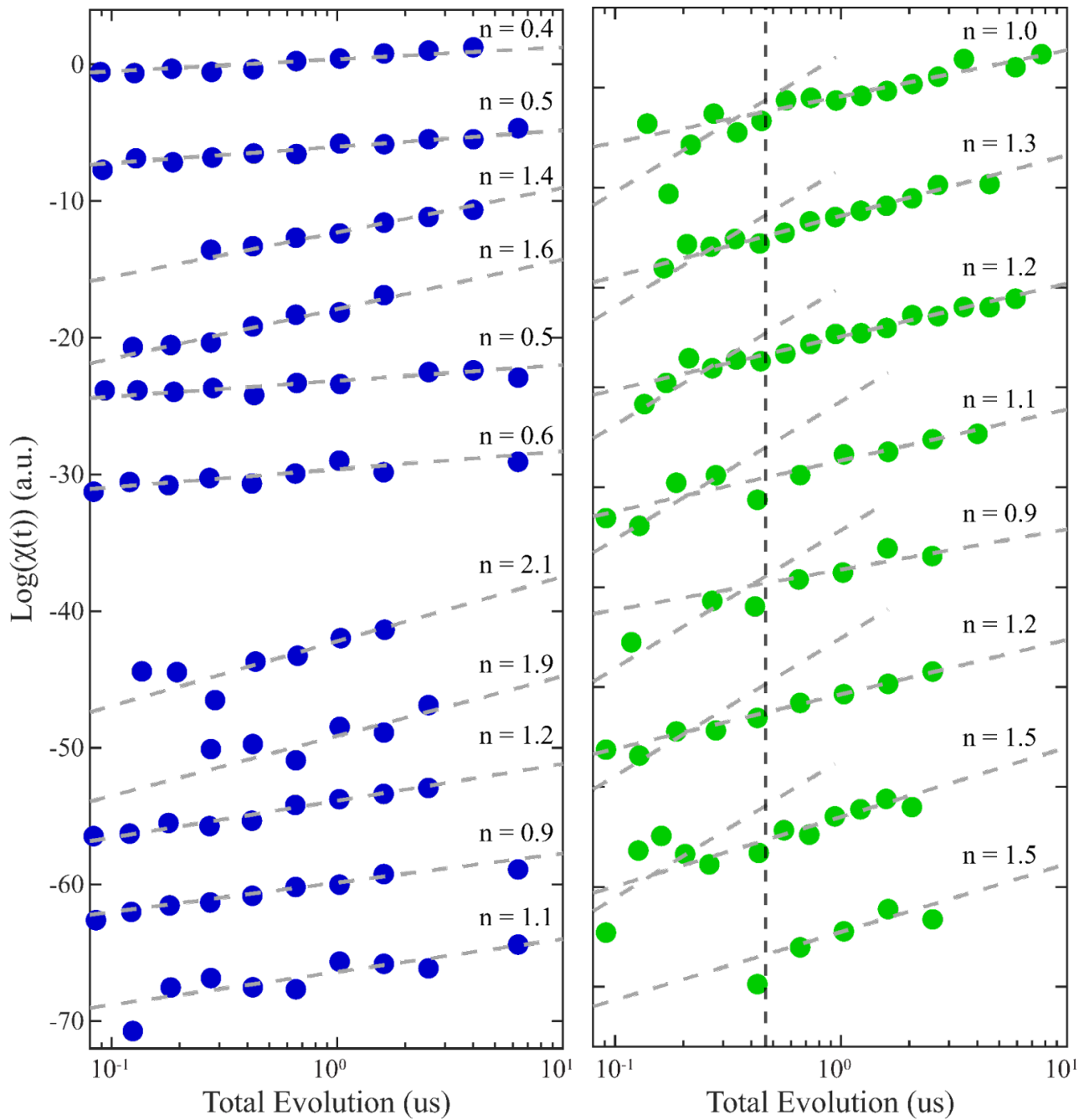

**Figure E10 — Echo stretch factors.** All bare (left panel) and core-shell (right panel) echo stretching factors were extracted as described in Figure 4C. Gray dashed lines are exponential fits for the random walk regime (the cutoff is marked by the black dashed line). For bare particles, the cutoff is too short to plot (30ns). Core-shell particles also show ballistic-regime fits to $n = 3$ to guide the eye (see methods and SI11 for more details about cutoff choice and plotting procedure).